\documentclass[10pt, conference, compsocconf]{IEEEtran}
\IEEEoverridecommandlockouts

\usepackage[utf8]{inputenc}
\usepackage[english]{babel}
\usepackage{graphicx}
\usepackage{xcolor}
\usepackage{cite}
\usepackage[T1]{fontenc} 
\usepackage{amsmath}
\usepackage{multirow}
\usepackage{bm}
\usepackage{array}
\usepackage{url}
\usepackage{psfrag}
\usepackage{mathtools}
\usepackage{amssymb}
\usepackage[linesnumbered,ruled]{algorithm2e}
\usepackage{algorithmic}

\usepackage{graphicx}
\usepackage{soul}
\usepackage[normalem]{ulem}

\usepackage{caption}
\usepackage{subcaption}

\SetKwInput{KwInput}{Input}                
\SetKwInput{KwOutput}{Output}              
\SetKwInput{kwInit}{Initialize}

\DeclareMathOperator{\EX}{\mathbb{E}}
\DeclareMathOperator*{\minimize}{minimize}

\DeclareMathOperator*{\diag}{diag}
\usepackage{soul}
\usepackage[linesnumbered,ruled]{algorithm2e}
\usepackage[acronym]{glossaries}
\usepackage{comment}

\usepackage{siunitx}
\usepackage{booktabs}

\usepackage{wrapfig}
\def\BibTeX{{\rm B\kern-.05em{\sc i\kern-.025em b}\kern-.08em
    T\kern-.1667em\lower.7ex\hbox{E}\kern-.125emX}}
\makeatletter
\def\blfootnote{\xdef\@thefnmark{}\@footnotetext}
\makeatother

\newacronym{iot}{IoT}{internet of things}
\newacronym{wncs}{WNCS}{Wireless Networked Control Systems}
\newacronym{ul}{UL}{uplink}
\newacronym{dl}{DL}{downlink}
\newacronym{rx}{RX}{receiver}
\newacronym{tx}{TX}{transmitter}
\newacronym{awgn}{AWGN}{additive white Gaussian noise}
\newacronym{snr}{SNR}{signal-to-noise ratio}
\newacronym{sdr}{SDR}{signal to distortion ratio}
\newacronym{mmse}{MMSE}{minimum mean square error}
\newacronym{gpr}{GPR}{Gaussian Process Regression}
\newacronym{gp}{GP}{Gaussian process}
\newacronym{lqr}{LQR}{Linear Quadratic Regulator}
\newacronym{bs}{BS}{base station}
\newacronym{aoi}{AoI}{Age-of-Information}
\newacronym{dare}{DARE}{discrete-time algebraic Riccati equation}
\newacronym{csi}{CSI}{channel state information}
\newacronym{rl}{RL}{Reinforcement Learning}
\newacronym{arima}{ARIMA}{Autoregressive Integrated Moving Average}
\newacronym{ml}{ML}{Machine Learning}
\newacronym{cdf}{CDF}{Cumulative Density Function}
\newacronym{mdp}{MDP}{Markov Decision Process}
\newacronym{rr}{RR}{Round Robin}
\newacronym{ca}{CA}{Control-Aware Scheduler}

\begin{document}

\clearpage

\title{
Resource Optimization for Tail-Based Control in Wireless Networked Control Systems 
}
\author{
\IEEEauthorblockN{Rasika~Vijithasena\IEEEauthorrefmark{1}\IEEEauthorrefmark{2},~\IEEEmembership{Member,~IEEE,} Rafaela~Scaciota\IEEEauthorrefmark{1}\IEEEauthorrefmark{2},~\IEEEmembership{Member,~IEEE,}  
Mehdi Bennis\IEEEauthorrefmark{1},~\IEEEmembership{Fellow,~IEEE}, \\and Sumudu~Samarakoon\IEEEauthorrefmark{1}\IEEEauthorrefmark{2},~\IEEEmembership{Member,~IEEE}
}
\IEEEauthorblockA{
	\small%
	\IEEEauthorrefmark{1}%
	Centre for Wireless Communication, University of Oulu, Finland \\	
	\IEEEauthorrefmark{2}%
	Infotech Oulu, University of Oulu, Finland 
     \\
     Email: \{rasika.vijithasena, rafaela.scaciotatimoesdasilva,  mehdi.bennis, sumudu.samarakoon\}@oulu.fi 
}}

\maketitle

\begin{abstract}

Achieving control stability is one of the key design challenges of scalable \gls{wncs} under limited communication and computing resources.
This paper explores the use of an alternative control concept defined as \emph{tail-based control}, which extends the classical \gls{lqr} cost function for multiple dynamic control systems over a shared wireless network. 
We cast the control of multiple control systems as a network-wide optimization problem and decouple it in terms of sensor scheduling, plant state prediction, and control policies.  
Toward this, we propose a solution consisting of a scheduling algorithm based on Lyapunov optimization for sensing, a mechanism based on \gls{gpr} for state prediction and uncertainty estimation, and a control policy based on \gls{rl} to ensure tail-based control stability.
A set of discrete time-invariant mountain car control systems is used to evaluate the proposed solution and is compared against four variants that use state-of-the-art scheduling, prediction, and control methods.
The experimental results indicate that the proposed method yields $22\%$ reduction in overall cost in terms of communication and control resource utilization compared to state-of-the-art methods.

\end{abstract}

\glsresetall

\section{Introduction}

\blfootnote{
This work is funded by the European Union Projects 6G-INTENSE (GA 101139266), VERGE (GA 101096034) and the project Infotech R2D2.
Views and opinions expressed are however those of the author(s) only and do not necessarily reflect those of the European Union. 
Neither the European Union nor the granting authority can be held responsible for them.
}

Advances in industrial control systems have spurred extensive research and innovation in wireless control systems in Industry 4.0~\cite{varghese2014wireless}. 
In particular, \gls{wncs}, defined as closed-loop control systems consisting of sensors, actuators, and controllers over wireless communication networks, are gaining prominence due to their cost-effectiveness in deployment and maintenance, adaptability in installations, and the potential to enhance safety. 
One of the significant challenges in the design of \gls{wncs} is the need for time-criticality that mandates seamless low-latency and reliability in wireless connectivity, which in turns impacts control stability~\cite{varghese2014wireless}.
In this view, resource allocation and scheduling of these control systems are crucial in attaining both latency and reliability targets.

The existing literature on \gls{wncs} focuses mainly on meeting latency and reliability through efficient resource allocation and scheduling~\cite{schenato2007foundations, xu2013stability, liu2003framework}.
 In~\cite{schenato2007foundations},
the authors utilize a \gls{rr} scheduling, which is an established static scheduling technique. 
In this approach, every sensor or controller transmits information consistently. 
This method cannot stabilize a large number of control systems due to the absence of scheduling decisions based on \gls{csi}.
To address this, a dynamic \gls{csi}-aware round-robin scheduling approach has been proposed in~\cite{xu2013stability, liu2003framework} so that energy consumption within the system is minimized while ensuring communication reliability.
Nonetheless, this method does not ensure stability for a substantial number of control systems, as the waiting time for these systems to refresh their state/action information increases with the number of served control systems.
As a remedy, the predictive control approach has been introduced in~\cite{girgis2021predictive} to ensure stability under scalability with limited communication resources.
Therefore, the freshness of the state information referred to as \gls{aoi} is considered for scheduling control systems, while \gls{gpr} is used to predict the states (and actions) for the non-scheduled sensors. 
To address the increased computation cost incurs in \gls{gpr}, the work of~\cite{ranasinghe2022predictive} opportunistically switches between high and low complexity prediction models to utilize resources while ensuring controllability at scale.
However, in both \cite{girgis2021predictive} and \cite{ranasinghe2022predictive}, no effort has been made to find alternative stability criteria that potentially reduce the total energy consumption.  

The main contribution of this paper is to propose a novel solution that improves the use of communication and control resources in \gls{wncs}.
Toward this end, we define a \emph{tail-based} stability objective that relaxes the strict stability conditions defined in the classical \gls{lqr} formulation.
Then the joint control and communication in the \gls{wncs} is carried out with three sub-tasks:  
i) scheduling a control system using the Lyapunov optimization framework,
ii) estimating the states of the unscheduled systems using \gls{gpr}, and
iii) deriving the best control policy, referred to as \emph{tail-based control}, hereinafter, over the proposed stability criteria using \gls{rl}.
%
In practice, the proposed solution is showcased through the path following of multiple autonomous robots. The robots navigate within bounded path areas rather than strictly adhering to a desired path. Action commands are only issued when the robots deviate from these boundaries. This approach improves cost efficiency by allowing for more relaxed controls within the specified boundaries.
The paper is organized as follows. 
Sec.~\ref{sec:system} presents the system model architecture.
The joint communication and control problem is formalized in Sec.~\ref{sec:problem} and the proposed solution is derived in Sec.~\ref{sec:proposed}.
In Sec.~\ref{sec:results}, numerical evaluations and results are discussed.
Finally, conclusions are drawn in Section~\ref{sec:concl}.

\emph{\textbf{Notation:}}
Scalars are denoted by lowercase symbols and sets are denoted by calligraphic letters.
Boldface lowercase letters denote column vectors, while boldface uppercase letters denote matrices. 
The $D \times D$ identity matrix is denoted by $\mathbf{I}_D$ and the spaces of $D \times D$ positive semi-definite, definite and positive definite matrices are denoted by $\mathbb{S}^{D}$, $\mathbb{S}^{D}_+$ and $\mathbb{S}^{D}_{++}$, respectively.
A multivariate Gaussian distribution with mean $\mathbf{m}$ and covariance $\mathbf{R}$ is denoted by $\mathcal{N}(\mathbf{m}, \mathbf{R})$ while $\EX[\cdot]$ is used for statistical expectation.
The operators $||\cdot||_p$, $(\cdot)^T$, and $\text{Tr}[.]$ stand for $p$-norm, transpose, and trace, respectively. 
The indicator function is defined as $\mathbb{I}_{\{a > b\}}=1$, if $a > b$ is held, and $\mathbb{I}_{\{a > b\}}=0$ otherwise.
 
\section{System Model}\label{sec:system}

\subsection{Wireless Networked Control Systems}

\begin{figure}[t]
\centering
\includegraphics[width=0.83\linewidth]{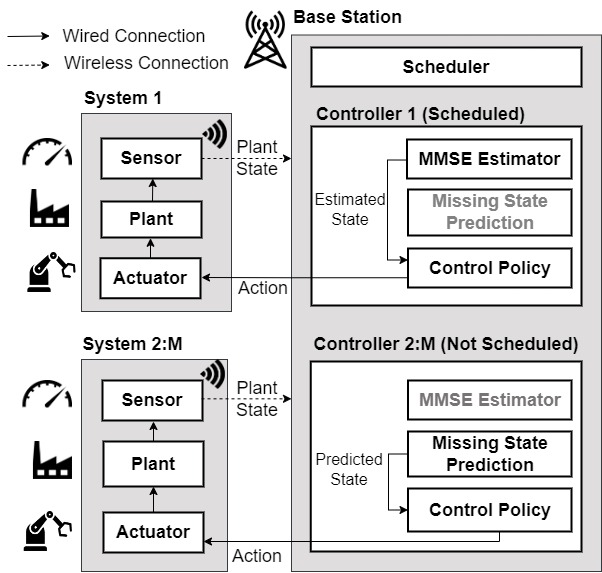} 
\caption{An illustration of $M$ number of \gls{wncs} where sensor-controller communication takes place over a shared wireless network.}
\label{fig1}
\end{figure}

We consider a \gls{wncs} setting which consists of a set $\mathcal{M}$ of $M$ non-linear control systems, in which, each system comprises a plant, sensor, controller, and an actuator along with a shared \gls{bs} as presented in Fig.~\ref{fig1}. 
At each time $k$, in the system $m\in\mathcal{M}$, the sensor observes the $D$-dimensional plant state $\mathbf{x}_{m,k}=[{x}_{m,k}^d]_{d\in\{1,\dots,D\}}\in\mathcal{X}_m$ where the state space of a system is $\mathcal{X}_m$. Then state is transmitted to the controller located at the \gls{bs} over a shared wireless channel.
Based on the received plant state, the controller computes the $N$-dimensional control action $\mathbf{{u}}_{m,k} \in \mathcal{U}$ where $\mathcal{U}$ is the action space and transmits to the actuator through a wired channel to stabilize the control system. 

The discrete-time dynamics of the system $m$ is 
\begin{equation} 
\label{dynamiceq}
\mathbf{{x}}_{m,k+1} = \mathbf{{A}}_{m} \mathbf{{x}}_{m,k} + \mathbf{B}_{m} \mathbf{u}_{m,k} + \mathbf{{w}}_{k},
\end{equation}
where $\mathbf{{A}}_{m} \in \mathbb{R}^{D \times D}$ and  $\mathbf{{B}}_{m} \in \mathbb{R}^{D \times N}$ are constant matrices that characterize the dynamics and $\mathbf{{w}}_{k} \in \mathbb{R}^{D}$ is the plant noise that is sampled from an independent and identically distributed (i.i.d) Gaussian distribution with zero mean and the covariance of $\mathbf{W}$. 
We assume that $\mathbf{{w}}_{k}$ contains the errors that occur due to the linearization of the non-linear control system.
Furthermore, $\mathbf{{A}}_{m}$ is assumed to be unstable if any eigenvalue has a magnitude greater than one~\cite{ogata1995discrete}, 
i.e. $|\lambda_{d}| > 1, $ where $ d \in \{ 1,...,{D}\}$, and $\lambda_{1},...,\lambda_{D}$ correspond to the eigenvalues of $\mathbf{{A}}_{m}$.
This suggests that, unless intervened, the plant's state will continuously increase (become unstable) over time.

\subsection{Wireless communication model}

The plant's state is transmitted to the controller located at the \gls{bs} from the sensor over a wireless communication channel, referred to as the \gls{ul}, which is presumed to be a Rayleigh block fading channel. 
This channel changes independently over time and remains static over time. 
In \gls{ul}, the received signal $\mathbf{y}_{m,k}$ in \gls{bs} for control system $m$ and time $k$ is
\begin{equation} 
 \label{eq10}
\mathbf{y}_{m,k}  = \sqrt{p_{m,k}} \mathbf{\Omega}_{m} \mathbf{H}_{m,k} \mathbf{x}_{m,k} + \mathbf{n}_{m,k},
 \end{equation}
where $p_{m,k} \in [ 0, P_{\text{max}}] $ is the transmission power of each control system with a maximum power $P_{\text{max}}$, $\mathbf{H}_{m,k} \in \mathbb{R}^{D \times D}$ is the wireless channel of the \gls{ul} communication of each control system, 
$\mathbf{\Omega}_{m} \in \mathbb{R}^{D \times D}$ is the observation matrix at the \gls{ul} state communication that characterizes the state-observation, 
and 
$\mathbf{n}_{m,k}$ is the additive white Gaussian noise with zero-mean and 
variance of $N_{0}$.
Thus, for a given bandwidth $\omega$, the \gls{ul} \gls{snr} is given by
 \begin{equation} 
 \label{eq11}
 \gamma_{m,k} = {p_{m,k} \Vert \mathbf{H}_{m,k} \Vert^{2}}/{N_{0}\omega }.
 \end{equation}  
For a successful \gls{ul} communication, $\gamma_{m,k} \geq \gamma_0$ needs to be satisfied with a given threshold $\gamma_0$.

\subsection{Scheduling UL Communication}

Since all controller units are co-located at the \gls{bs} and share the \gls{ul} communication channel, a scheduler deployed at the \gls{bs} is utilized to enable interference-free \gls{ul} communication. 
Towards this, the \gls{bs} uses a scheduling indicator $a_{m,k}$ where $a_{m,k}=1$ represents the channel is assigned for the \gls{ul} of system $m$ at time $k$ and $a_{m,k}=0$ otherwise.
When $a_{m,k}  = 1$, the \gls{mmse} estimator is used to decode the received signal as
\begin{equation} 
 \label{mmse_estimate}
 \tilde{\mathbf{x}}_{m,k}=\mathbb{E} [ \mathbf{x}_{m,k} | \mathbf{y}_{m,k} ] = \mathbf{G}_{m,k} \mathbf{y}_{m,k} = \mathbf{x}_{m,k} + \mathbf{v}_{m,k},  
 \end{equation}
where $\mathbf{G}_{m,k} = (\sqrt{p_{m,k}} \boldsymbol{S}_{m,k} \mathbf{H}^{T}_{m,k} \mathbf{\Omega}_{m}^{T})/( p_{m,k} \mathbf{\Omega}_{m} \mathbf{H}_{m,k} \boldsymbol{S}_{m,k} \\\mathbf{H}_{m,k}^{T} \mathbf{\Omega}_{m}^{T} + N_{0} \mathbf{I}_{D})$ is the linear \gls{mmse} matrix of control system $m$ and 
$\boldsymbol{S}_{m,k} = \mathbb{E}[\mathbf{x}_{k} \mathbf{x}_{k}^T]$ is the covariance matrix. 
Here, $\mathbf{v}_{m,k}$ is the \gls{mmse} estimation error that is assumed to be a Gaussian random vector with zero mean and $\mathbf{V}_{m,k}$ covariance matrix given by
\begin{equation}
 \label{mmse_error_cov}
 \mathbf{V}_{m,k} = \mathbb{E}[ \mathbf{v}_{m,k}\mathbf{v}_{m,k}^T ] = \boldsymbol{S}_{m,k} - \mathbf{G}_{m,k} \sqrt{p_{m,k}} \mathbf{\Omega}_{m} \mathbf{H}_{m,k} \boldsymbol{S}_{m,k}.
 \end{equation}

On other hand, if $a_{m,k}  = 0$, the controller $m$ relies on an estimate $\hat{\mathbf{x}}_{m,k}$ of the state to compute the control action. While opportunistically scheduling a control system, it is highly likely that control decisions derived from estimated and outdated states may become inaccurate, in which, ensuring stability could be challenging.  
In this view, the \gls{aoi} can be used to measure the freshness of the states considering the time interval between two consecutive state information updates. 
\Gls{aoi} linearly increases with time and can be represented as
\begin{equation}
    \beta_{m,k} =
    \begin{cases}
        1, & \text{if~} a_{m,k} = 1 \text{~and~} \gamma_{m,k}\geq\gamma_0, \\
        1 + \beta_{m,k-1}, & \text{otherwise}.
    \end{cases}
\end{equation}

The controller computes the control action based on the estimated state or predicted state considering \gls{aoi} and transmits it to the actuator. We consider \gls{dl} to be ideal, leading to the assumption that the actuator receives control decisions without any errors. Thereafter, the actuator applies these error-free control decisions on the plant to maintain stability.

\section{Joint Communication and Control Problem} \label{sec:problem}

Our objective of this research is to optimize the resources utilized for communication and control while ensuring the control stability and communication reliability.

\subsection{Communication Cost}
\label{subsec:problem communication cost}

The increase in \gls{aoi} reflects fewer communications, which is beneficial in terms of energy savings, but has an adverse effect on the estimated state with increased uncertainty. 
Since it can cause the system to be unstable, it is essential to strike a balance between \gls{aoi} measures over all systems.
At the same time, transmission power impacts the reliability of communication and the amount of energy consumed, and thus, power allocation for communication with all systems should be treated fairly. 
In this view, the total communication cost ${C}$ can be expressed as follows:
\begin{equation}
\label{comm_cost} 
{C} \left( [ \bar{\beta}_{m} ,  \bar{\hat{p}}_{m} ]_{m\in\mathcal{M}} \right) = \psi_{\beta} \sum_{m=1}^{M} \mathcal{G}( \bar{\beta}_{m})\\
+    \psi_{p} \sum_{m=1}^{M} \mathcal{G}( \bar{\hat{p}}_{m}) ,
\end{equation} 
where $\hat{p}_{m,k} = a_{m,k} {p}_{m,k}$ and $\mathcal{G}(.) = \log(1+.)$ is a non-decreasing concave function that captures the fairness of \gls{aoi} and transmission power for each control system~\cite{li2008proportional}.
Hereinafter, the notation of
$ \bar{z} = \frac{1}{K} \sum_{k=1}^{K} z_k $
is used for the time average value of any quantity $z_k$.
The positive coefficients $\psi_{\beta}$ and $\psi_{p}$ can be defined to reflect the relative importance of the costs related to \gls{aoi} and the transmission power, respectively.

\subsection{Control Cost}
\label{subsec:problem control cost}

Traditional \gls{lqr}-based control designs are rooted on the stability defined over a singular point of operation and thus, the optimal control always ensures that the plant state is always maintained around the desired state.
This design is inefficient for systems that consider extreme state conditions (tail conditions) as undesirable. 
Towards this, the objective of tail-based control is to define a state range for the desired stability, in contrast to the classical stability conditions.

In this view, for any system $m$, suppose that the desired range of stability $\|\mathbf{x}_{m,k}\|_p > \eta$ at time $k$ is defined by the threshold $\eta(>0)$.
Hence, the deviations from stability and the effort spent on control is used to define the control cost $J_{m,k}$ for a given control system $m$ at time $k$ as follows:
\begin{equation}
\label{costtail}
{J}_{m,k} =  \underbrace{ \mathbf{x}^T_{m,k} \mathbf{Q} \mathbf{x}_{m,k}  f_{m,k} }_\text{Cost related to stability}  + \underbrace{ \mathbf{u}^T_{m,k} \mathbf{Y} \mathbf{u}_{m,k} }_\text{Cost related to controlling} ,
\end{equation}
where the indicator $f_{m,k}=\mathbb{I}_{\{ \|\mathbf{x}_{m,k}\|_p > \eta \}}$.
%
Here, $\mathbf{Q}\in \mathbb{S}^{D}$ and $\mathbf{Y} \in \mathbb{R}$, which correspond to control system design parameters.

\subsection{Control-Constrained Optimization Problem}
\label{subsec:problem optimization problem}
 
We use the quadratic Lyapunov function to evaluate control stability, which measures the performance of each control system as a function of the state as follows:
\begin{equation}
\label{Lyapunov_fun}    
\mathcal{L}(\mathbf{x}_{m,k}) = \mathbf{x}^{T}_{m,k} \, \mathbf{Z} \, \mathbf{x}_{m,k} f_{m,k}, 
\end{equation}
where $\mathbf{Z} \in \mathbb{S}^{D}_{++}$ is a unique solution to the discrete Lyapunov equation 
$\mathbf{{A}}^{c^{T}}_{m} \mathbf{Z} + \mathbf{Z} \mathbf{A}^{c}_{m} = - \mathbf{I}_{D}$, and $\mathbf{{A}}^{c}_{m}$ is a closed-loop state transition matrix defined as $\mathbf{A}^{c}_{m} = \mathbf{A}_{m} - {\mathbf{B}}_{m} \mathbf{\Phi}_{m} $.
Here, $\mathbf{\Phi}_{m}$ is the feedback gain matrix of the control system $m$, which is given by,
\begin{equation}
\label{feedback_gain} 
\mathbf{\Phi}_{m}
= {\arg\min}_{\mathbf{\Phi}'_{m}}  \frac{1}{K} \sum_{k=1}^{K}(\mathbf{u}_{m,k} - \mathbf{\Phi}'_{m}\mathbf{x}_{m,k}).
\end{equation}
However, the centralized scheduler only has access to the predicted state $\hat{\mathbf{x}}_{m,k}$. Therefore, the expected Lyapunov value is calculated as~\cite{girgis2021predictive},
 \begin{equation}
 \label{Lyapunov_Expec} 
 \mathbb{E}\left[ \mathcal{L}(\mathbf{x}_{m,k}) \vert \hat{\mathbf{x}}_{m,k} \right]  = 
 \hat{\mathbf{x}}^{T}_{m,k} \, \mathbf{Z} \, \hat{\mathbf{x}}_{m,k}
 + \text{Tr} \left[  \mathbf{Z} \, \mathbf{\Psi}_{m,k} \right], 
 \end{equation}
where $\mathbf{\Psi}_{m,k}$ is the covariance matrix of the state prediction error, which is discussed in~\ref{Subsec:gpr_prediction}.
For control stability, it is required that the expected future value of the Lyapunov function $\mathcal{L}(\mathbf{x}_{m,k+1})$, should be less than the current value of $\mathcal{L}(\mathbf{x}_{m,k})$ scaled by a certain rate $\zeta_{m} \in (0, 1 ]$ as
\begin{multline}
\label{Future_Lyapunov}    
\mathbb{E}\left[ \mathcal{L}(\mathbf{x}_{m,k+1}) \vert \hat{\mathbf{x}}_{m,k}, \mathbf{u}_{m,k}, \mathbf{H}_{m,k}, p_{m,k}\right] \\
\leq \zeta_{m}  \mathbb{E}\left[ \mathcal{L}(\mathbf{x}_{m,k}) \vert \hat{\mathbf{x}}_{m,k} \right] ,
\end{multline} 
where the expectation in right hand side of~\eqref{Future_Lyapunov} is with respect to the plant noise $\mathbf{w}_k$ in~\eqref{dynamiceq} and the signal estimation error $\mathbf{v}_{m,k}$ defined in~\eqref{mmse_estimate}.

Based on the aforementioned costs of communication and control along the stability and \gls{ul} channel sharing (one system is scheduled at a time), the formulation of the control-constrained optimization problem is given by,
\begin{subequations}\label{main_cost_function} 
\begin{alignat}{1}
\underset{ (\mathbf{a}_{k}, \mathbf{p}_{k}, \mathbf{u}_{k}), \forall k  }{\minimize}  
&\quad
\frac{1}{M} \sum_{m=1}^{M} {C} \left(  \bar{\beta}_{m} ,  \bar{\hat{p}}_{m}  \right) + \frac{1}{M} \sum_{m=1}^{M}\bar{J}_{m}  \label{eqopt1e_1}\\
\text{s.t.}  
\label{eqopt1e_1a}
&\quad
	\gamma_{m,k} \geq a_{m,k} \gamma_0, \quad \forall m\in\mathcal{M}, \forall k,\\
        \label{eqopt1e_1d}
        &\quad \mathbf{a}_{k} \in \mathcal{A}, \mathbf{p}_{k} \in \mathcal {P},  \mathbf{u}_{k}\in \mathcal{U}, \quad \forall k, \\
        \label{eqopt1e_1e}
        &\quad \eqref{Future_Lyapunov}, \quad \forall m\in\mathcal{M},\forall k,
\end{alignat}
\end{subequations}
where $\mathcal{A} = \{\mathbf{a}_{k} = [a_{m,k}]_{m \in \mathcal{M}} | \sum_m a_{m,k} \leq 1, a_{m,k}\in\{0,1\}\}$
and
$\mathcal{P} = \{\mathbf{p}_{k} = [p_{m,k}]_{m \in \mathcal{M}} | p_{m,k} \in [0,P_{\text{max}} \}$.

\section{Communication and Control Co-Design} \label{sec:proposed}

Due to the complex nature of the joint control and communication problem posed in~\eqref{main_cost_function}, we propose a solution method consisting of three components:
an \gls{ul} scheduling algorithm, a \gls{gpr}-based prediction model for state prediction, and a control policy to satisfy the tail-based stability.

\subsection{Lyapunov Optimization for UL Scheduling} \label{Subsec:lyapunov}

In the stability constraint~\eqref{Future_Lyapunov}, the decisions related to state prediction, UL scheduling, and tail-based control are coupled, in which, solving~\eqref{main_cost_function} becomes a daunting task. 
Hence, ~\eqref{Future_Lyapunov} can be recast in an alternative form as follows~\cite[Sec.~IIIA]{girgis2021predictive}:
\begin{equation}\label{lyapunov_second}
{\lim\sup_{K\to\infty}} \frac{1}{K} \sum_{k=1}^{K} \xi_{m,k} \geq  
{\lim\sup_{K\to\infty}} \frac{1}{K} \sum_{k=1}^{K} \frac{\Gamma_{m,k}}{\Upsilon_{m,k}}
\end{equation}
where $\xi_{m,k} = a_{m,k}\mathbb{I}_{\{\gamma_{m,k} > \gamma_0\}}$, and 
\begin{align*}
    \Upsilon_{m,k}=
    &\begin{multlined}[t]
    \mathrm{Tr}[(\mathbf{B}_{m} \mathbf{\Phi}_{m})^{T} \mathbf{Z} (\mathbf{B}_{m} \mathbf{\Phi}_{m})\mathbf{\Psi}_{m,k}] 
    \\  
    - \mathrm{Tr}[ (\mathbf{B}_{m} \mathbf{\Phi}_{m})^{T} \mathbf{Z} (\mathbf{B}_{m} \mathbf{\Phi}_{m}) \mathbf{V}_{m,k} ], 
    \end{multlined}
    \\
    \Gamma_{m,k} =
    &\begin{multlined}[t]
    \left\| (\mathbf{A}_{m}^{c} - \zeta_{m}\mathbf{I}_D)_{m,k} \right\|^{2}_{\mathbf{Z}^{\frac{1}{2}}} 
    + \mathrm{Tr}[\mathbf{B}_{m}^T \mathbf{Z} \mathbf{B}_{m}\mathbf{\Psi}_{m,k} ]
    \\
    + \mathrm{Tr}\left[(\mathbf{A}_{m}^{T} \mathbf{Z} \mathbf{A}_{m} - \zeta_{m} \mathbf{Z})\mathbf{\Psi}_{m,k}\right]  
    + \mathrm{Tr}[\mathbf{Z}\mathbf{W}]. 
    \end{multlined} 
\end{align*}

Due to the decoupling of scheduling, prediction, and control variables as~\eqref{lyapunov_second}, we can now minimize the communication cost $C$ independently given the control policy as follows:
%
\begin{subequations}\label{second_cost_problem} 
\begin{eqnarray}
&\underset{(\mathbf{a}_{k}, \mathbf{p}_{k}), \forall k}{\minimize} \quad 
& \frac{1}{M} \sum_{m=1}^{M}{C} \left( \bar{\beta}_{m} ,  \bar{\hat{p}}_{m}  \right) \label{eqopt1e_2}
\\
&\text{s.t.} 
\label{eqopt1e_2a}
& \bar{a}_{m} \geq \bar{{G}_{l}}( {c}_{m,k}), \quad\forall m\in\mathcal{M},\\
\label{eqopt1e_2b}
&& \eqref{eqopt1e_1a},~\eqref{eqopt1e_1d},
\end{eqnarray}
\end{subequations}
where ${c}_{m,k} = \Gamma_{m,k}/\Upsilon_{m,k}$ is the lower-bound stability of $\xi_{m,k}$ of each control system. 
Here, ${G}_{l}(\cdot) = \max \left[ \min (\cdot, 1), 0 \right] ]$ is used as the lower bound of \eqref{eqopt1e_2a}
to ensure the feasibility of the scheduling constraints $\sum_{m=1}^M a_{m,k} \leq 1$ and $a_{m,k} \in \{0,1\}$ for all $k$. 
To solve the optimization over a time horizon, we resort to the stochastic Lyapunov optimization framework \cite{neely2010queue}.
Therein, the notion of virtual queues is used to track the dynamics of the time average variables, allowing us to decompose~\eqref{second_cost_problem} into a series of sub-problems that can be solved at each time step $k$.
The details of the solution are similar to the work of~\cite{girgis2021predictive} and left out due to the space limitations.
We direct the interested readers to~\cite[Sec.IV]{girgis2021predictive} for the detailed derivations.


\subsection{Prediction of Missing States using \gls{gpr}} \label{Subsec:gpr_prediction}

In the absence of \gls{ul}, either due to poor \gls{snr} or not being scheduled, controllers are required to predict the state of their corresponding systems using previous observations. 
For this, we employ a \gls{gpr}-based state prediction technique in each controller.
Since the system state dynamics results in a time series, the dynamics can be modeled as sampling from a Gaussian process.

Let $\mathbf{X}_{m,\boldsymbol{k'}}^d = [\tilde{x}_{m,k'}^d]_{k'\in\mathcal{K}}$ be past observations
where $\mathcal{K}=\{ \kappa|\kappa<k, a_{m,\kappa}=1, \gamma_{m.\kappa}\geq\gamma_0 \}$ with the corresponding time steps $\boldsymbol{k}'=[k']_{k'\in\mathcal{K}}$.
Then the latent function of the regression model over the dimension $d$ can be learned as $\tilde{x}_{m,k'}^d = g_d(k') + \tau$ with a noise $\tau \sim \mathcal{N} \left( {0} , \sigma^2_{\kappa} \right)$.
For estimating $g_d$, first, it is necessary to rely on a correlation between the state observations over different time steps. 
In this view, we consider a periodic kernel to model the correlation between the outputs according to their temporal behaviors as~\cite{rasmussen2006gaussian},
\begin{equation}  
 \label{eq16}
 {R}(k,k') =  
 \textstyle 
 h^{2} \exp\left( -\frac{2}{l^2} \sin^{2} \left( \frac{\pi \left(k - k' \right)}{s} \right) \right),
\end{equation}
where ${R}(k,k')$ represents the covariance between the outputs according to times $k$ and $k'$.
Here, $h$, $l$, $s$ are the hyper-parameters that represent output scaling, time scaling, and frequency, respectively, which are denoted as $\mathbf{\Theta}=(h, l, s)$, hereinafter.
Then, the joint distribution of past observations and the output of $g_d(k)$ follows:
\begin{equation}
 \label{eq17}
 \left[ \begin{array}{c}
     \mathbf{X}^d_{m,\boldsymbol{k}'}  \\
     g_d(k)
 \end{array}  \right]  \sim \mathcal{N} \left( 
     \mathbf{0} , 
     \left[ \begin{array}{cc}
     \mathbf{R}(\boldsymbol{k}') &  \mathbf{r}(\boldsymbol{k}',k) \\
     \mathbf{r}(k,\boldsymbol{k}') & \mathbf{r}(k,k)
 \end{array}  \right]  \right),
\end{equation} 
where $\mathbf{R}(\boldsymbol{k}')=[R(i,j)]_{i,j\in\mathcal{K}}$, $\mathbf{r}(k,\boldsymbol{k}') = [R(k,j)]_{j\in\mathcal{K}}$, $\mathbf{r}(\boldsymbol{k}',k) = \mathbf{r}(k,\boldsymbol{k}')^T$ and $K=|\mathcal{K}|$.
Then, the posterior distribution of $g_d(k)$ based on the past observations $\mathbf{X}^d_{m,\boldsymbol{k'}}$ can be analytically determined as,
\begin{equation}
\label{eq18}
\text{Pr}\left( g_d(k)  |   \mathbf{X}^d_{m,\boldsymbol{k'}}, k, {\mathbf{\Theta}} \right) \sim \mathcal{N} \left( \hat{x}^d_{m,k}, (\sigma^{d}_{m,k})^2 \right).
\end{equation} 
%
Here, $\sigma_{m,k}^d = \sqrt{( {R}(k,k) - \mathbf{r}(k,\boldsymbol{k}')\mathbf{R}(\boldsymbol{k}')^{-1} \mathbf{r}(k,\boldsymbol{k}')^T )}$ is the standard deviation and the predicted state over the dimension $d$ is given by
\begin{equation}
\hat{x}_{m,k}^d = \mathbf{r}( k,\boldsymbol{k}')^T\mathbf{R}(\boldsymbol{k}')^{-1}\mathbf{X}^d_{m,\boldsymbol{k}'}. \label{gpr_pred_mean}    
\end{equation}

Finally, the prediction error covariance matrix is calculated as 
\begin{equation}\label{predicted_covariance}
\mathbf{\Psi}_{m,k} = \diag \left( \sigma_{m,k}^2(1), \dots, \sigma^2_{m,k}(D) \right).
\end{equation}

\subsection{Tail-based Control}\label{Subsec:tail_control}

The control cost minimization can be represented as a series of sub-problems, in which, for system $m$ it is as follows: 
%
\begin{align}\label{control_cost_eq} 
\minimize_{ (\mathbf{u}_{m,k} \in \mathcal{U}), \forall k  } \quad &  
\frac{1}{M} \sum_{m=1}^{M} \bar{J}_{m}.
\end{align}
%

Due to the inherent complexities and potential non-linearities of tail-based control approach, it is a challenge to find a close form expression for the control policy by using either convex optimization, dynamic programming, or numerical methods.  
In contrast, it is efficient at optimizing a non-linear, complex cost function by refining the policy in an iterative manner through trial and error. 
Hence, the derivation of the optimal control policy $\boldsymbol{\pi}^*$, the probabilistic mapping between plant states and control actions, has been cast as a solving \gls{mdp} using \gls{rl}. 
The \gls{mdp} problem is formulated as a tuple $(\mathcal{X}_m, \mathcal{U}, F, {R_e}, \nu)$ where $F$ defines the dynamics as per \eqref{dynamiceq} and $\nu$ is the discount factor. 
Considering the tail-based control cost $J_m$ in~\eqref{costtail}, 
The reward ${R_e}(\mathbf{x}_{m,k},\mathbf{u}_{m,k})$ is generated from the tail-based control cost $J_{m,k}$ in~\eqref{costtail} as follows:
\begin{equation}
\label{rlreward} 
{R_e}(\mathbf{x}_{m,k},\mathbf{u}_{m,k}) =  -J_{m,k}.
\end{equation}
In the above \gls{mdp}, the optimal policy $\boldsymbol{\pi}^*$, which maximizes the discounted cumulative rewards can be represented as 
\begin{equation}
\label{rloptimalpolicy}
\boldsymbol{\pi}^* = \max_{\boldsymbol{\pi}} \left[ \sum_{k=1}^\infty \nu^k {R_e}(\mathbf{x}_{m,k},\mathbf{u}_{m,k} | \boldsymbol{\pi} ) \right].
\end{equation}

Afterwards, the optimal control decision is derived by sampling the control policy given the state, i.e., $\mathbf{u}_{m,k}={B}_u(\mathbf{x}_{m,k};
\boldsymbol{\pi})$ where $B_u$ represents the function in our \gls{rl} model. It maps the current state of the system, $\mathbf{x}_{m,k}$, to an optimal control action, $\mathbf{u}_{m,k}$, using the policy $\boldsymbol{\pi}$ that has been learned through the \gls{rl} process.

\subsection{Proposed Algorithm}

Due to the decoupling of UL scheduling, state prediction, and control policy. we implement the overall solution in two-step. In the first step, the RL-based policy~\eqref{rloptimalpolicy} is trained over the reward function~\eqref{rlreward} for a generic system. 
During the second step, the feedback gain matrix is computed using~\eqref{feedback_gain} and used for UL scheduling followed by state prediction. Finally, the control decisions are computed using either received or estimated state. These steps are presented in Algorithm~\ref{alg:tail-based}.

\begin{algorithm}[!b]
\caption{Communication and Control Co-Design}
\begin{algorithmic} [1]
\STATE \textbf{Step 1: Offline training of the tail-based control policy}
\STATE Initialization: $\mathcal{X}, \mathcal{U}, \nu$, $\mathbf{x}_{0}$, $\mathbb{\eta}$, and $\boldsymbol{\pi}$
\FOR{each epoch} 
\FOR{each step $k$} 
\STATE Sample action $\mathbf{u}_{k}={B}(\mathbf{x}_{k}; \boldsymbol{\pi})$ 
\STATE Calculate reward ${R_e}$ using~\eqref{rlreward}
\STATE Update state $\mathbf{x}_{k+1}$ using~\eqref{dynamiceq}
\STATE Update policy $\boldsymbol{\pi}$ using~\eqref{rloptimalpolicy} 
\ENDFOR
\ENDFOR

\item \textbf{Step 2: Online decision making at the controller }
\STATE Initialization: $V, \psi_\beta, \psi_p$, and $ \mathbf{x}_{m,0}$
\FOR{$k$ =1 to $K$}
\FOR{${m}$ = 1 to $M$}
\STATE Compute the feedback gain matrix $\mathbf{\Phi}$ by~\eqref{feedback_gain}
\STATE Calculate $\hat{\mathbf{x}}_{m,k}$ with~\eqref{gpr_pred_mean} and $\mathbf{\Psi}_{m,k}$ using~\eqref{predicted_covariance}
\STATE Find $a_{m,k}$ by solving~\eqref{second_cost_problem}
\IF{$a_{m,k} = 1$}
\STATE Schedule system $m$ and decode $\tilde{\mathbf{x}}_{m,k}$ using \eqref{mmse_estimate}
\ELSE
\STATE Predict $\hat{\mathbf{x}}_{m,k}$ using~\eqref{gpr_pred_mean}
\ENDIF
\STATE Apply control action $\mathbf{u}_{m,k}={B_u}(\mathbf{x}_{m,k};
\boldsymbol{\pi})$
\STATE Update $\mathbf{x}_{m,k+1}$ based on $\mathbf{u}_{m,k}$ using~\eqref{dynamiceq} 
\ENDFOR
\ENDFOR
\end{algorithmic}\label{alg:tail-based}
\end{algorithm}

\newcommand{\full}{\texttt{Full}}
\newcommand{\vOne}{\texttt{V1}}
\newcommand{\vTwo}{\texttt{V2}}
\newcommand{\vThree}{\texttt{V3}}
\newcommand{\vFour}{\texttt{V4}}

\section{Numerical Results}\label{sec:results}

In this section, we present numerical results to validate our theoretical analysis. We employ the Moore’s mountain car problem~\cite{moore1990efficient} to evaluate the performance of the proposed algorithm.  
The discretized state space dynamics that use scalar control actions with a sampling period $T = 0.01$\,s and the equilibrium point $ \epsilon =[\frac{\pi}{2b}, 0]$ results in
\begin{equation}
\mathbf{A}_m = 
\begin{bmatrix}
    (1+\alpha{b}) & 1 \\
    \alpha{b} & 1
\end{bmatrix},
\quad 
\mathbf{B}_m = \begin{bmatrix}
1 \\
1 \\
\end{bmatrix},
\quad \text{and} \quad
\mathbf{Y} = 1,
\label{eq:sistema}
\end{equation}
with $\alpha$ as the gravity. 
We use the following as the default simulation parameters $\mathbf{Q} = \mathbf{I}$, $\mathbf{x}_{m,0} = [-1.5, 0]$, $b=3$, $\alpha=0.0025$\,Nkg\textsuperscript{-1}, $\mathbf{\Omega}= \mathbf{I} $, $P_{max} = 30$\,dBm, $N_0 = 1$,  $V = 1000$, $\zeta_i = 0.1$, $ \psi_{\beta} = 1$, $\psi_{p} = 1$, $\gamma_{0} = 20$\,dB, $\sigma_m^2 = 0.02$, and $p_{m,k} = 28.2$\,dBm unless specified otherwise.
For the \gls{gpr}, $10$ past state observations have been used. 
For each simulation, $1000$ time steps have been carried out.
Furthermore, all analyses are based on averaging data collected over $100$ simulations.

The \gls{rl} model is developed using the Python MushroomRL library~\cite{mushroomrl}, which provides common interfaces to develop \gls{rl} models. 
A Gaussian policy with a learnable standard deviation has been considered to train the policy. The \gls{rl} model is trained for $100$ epochs, each with $200$ episodes with $\eta = [0.1, 0.1]$, $\mathcal{X} =  [-\infty, +\infty], \mathcal{U} = [-10, 10] $, and $ \nu = 0.9$.

Since the proposed solution consists of three parts: a novel \gls{ca}, the adoption of \gls{gpr} and a RL-based control policy, the corresponding concepts of the existing solutions, the \gls{rr} scheduler, without predictions or \gls{arima}~\cite{brockwell2002introduction}, and the classical \gls{lqr} control policy, respectively, are used for performance comparison.
In this regard, we compare the proposed solution \full{} with four other variants summarized in Table~\ref{tab:methods}.
\begin{table}[!b]
    \caption{Methods compared in simulations}
    \label{tab:methods}
    \centering
        \begin{tabular}{@{}c ccc@{} l@{}}
            \cmidrule{1-4}
            \textbf{Method} & \textbf{Scheduling} & \textbf{State prediction} & \textbf{Control}\\
            \cmidrule{1-4}
            \vOne & RR & No &Tail& \multirow{4}{*}{\hspace{-1em}$\left.\begin{array}{l}
                \\
                \\
                \\
                \\
                \end{array}\right\rbrace\text{Variants}$} \\
            \vTwo & RR & ARIMA &Tail& \\
            \vThree & RR & ARIMA &Classic& \\
            \vFour & CA & GPR &Classic&\\
            \full & CA & GPR &Tail\\
            \cmidrule{1-4}
        \end{tabular}
\end{table}
\begin{figure}[!t]
    \centering
    \includegraphics[width=0.5\textwidth]{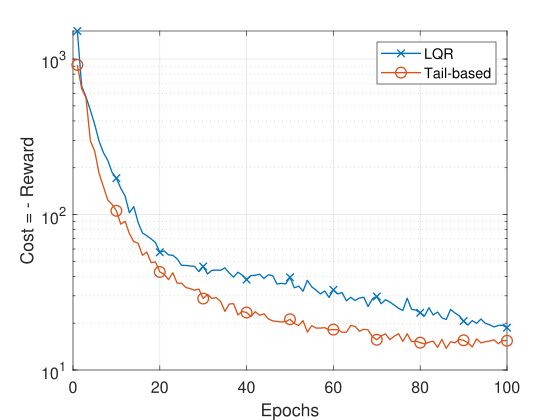}
    \caption{RL convergence of the tail-based control and LQR problems.}
    \label{fig:rl_policy}
\end{figure}

First, we analyze the convergence of the trained \gls{rl} policy via simulations for both the proposed tail-based control and \gls{lqr} formulations as illustrated in Fig.~\ref{fig:rl_policy}. 
Though there exists a closed form solution for \gls{lqr} formulation, we have used \gls{rl} for deriving both policies for a fair comparison. 
It can be noted from Fig.~\ref{fig:rl_policy} that both policies achieve convergence with respective to their own objectives after about $70$ epochs.

\begin{figure}
      \centering
        \begin{subfigure}{0.5\textwidth}
            \includegraphics[width=\textwidth]{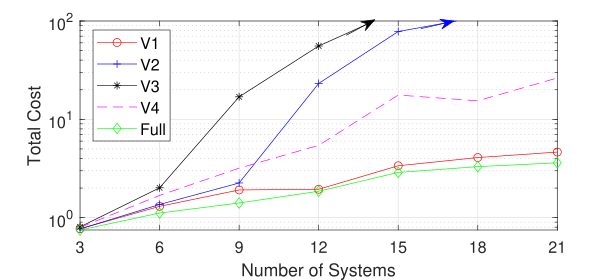}
            \caption{Objective value as per \eqref{eqopt1e_1}.}
            \label{fig:total_cost}
        \end{subfigure}
	\vfill
	          \centering
     \begin{subfigure}[t]{0.48\linewidth}
         \centering
         \includegraphics[width=\linewidth, keepaspectratio]{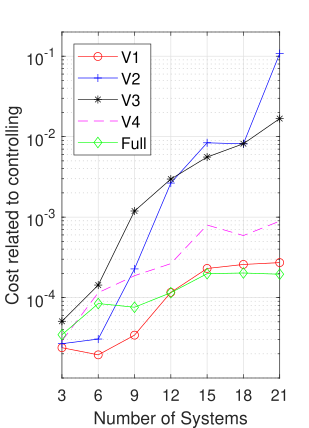}
         \caption{Cost related to controlling.}
         \label{fig:control_cost}
     \end{subfigure}
     \hfill
     \begin{subfigure}[t]{0.48\linewidth}
         \centering
         \includegraphics[width=\linewidth]{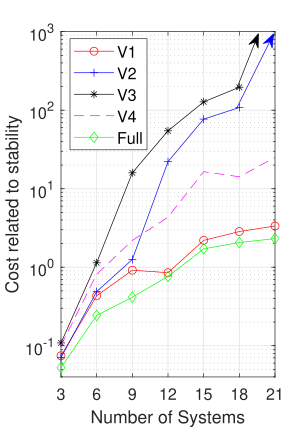}
         \caption{Cost related to stability.}
         \label{fig:state_cost}
     \end{subfigure}
    \caption{Performance of total cost, cost related to controlling, and cost related to stability per control system for different $M$.}
    \label{fig:two_graphs}
    \vspace{-0.6em}
\end{figure}

In Fig.~\ref{fig:two_graphs}, we compare the total cost that is the combination of communication cost~\eqref{comm_cost} and control cost~\eqref{costtail}, the cost related to controlling, and the cost related to stability for all methods with settings having different number of control systems. 
Specifically, in Fig.~\ref{fig:total_cost}, we can observe that \vOne{}, exhibits lower costs compared to other three variants while the proposed solution \full{} yields the lowest cost.
In summary, the reductions of total cost with \full{} compared to \vOne{} are $15\%$ at $M=6$ while $22\%$ at $M=21$. 
%

%

Fig.~\ref{fig:control_cost} shows that for $M=6$, \vOne{} and \vTwo{} have costs related to controlling $4.3$ and $2.8$ times lower than the \full{} method, respectively. However, as the number of systems scales to $M=21$, \vTwo{} incurs higher costs, while \vOne{} exhibits slightly higher costs than \full{}.
This is due to the fact that the \gls{arima} method tends to produce inaccurate state estimates leading to system instability as well as to apply control decisions with higher magnitude. 
On the contrary, \vOne{}, which lacks a prediction method, avoids applying control decisions when systems are unscheduled, thus, avoiding the undesirable state fluctuations evident in \vTwo{}. 
%
%
Meanwhile, the classic control policy employed in \vThree{} and \vFour{} is focused on imposing strict stability, which is impractical when control decisions are applied inconsistently, thus, yielding increased costs.
%
Furthermore, we can observe in Fig.~\ref{fig:state_cost}, the costs related to stability are comparatively much higher (about $10^6$ times) than the costs related to controlling, which are dominant in the overall cost computation.
As a result, the total costs in Fig.~\ref{fig:total_cost} and the costs related to stability in Fig.~\ref{fig:state_cost} show similar trends. Finally, we would like to point out that certain instances showcase fluctuations in cost that disagree with trends, which can be rectified by augmenting the number of simulations.

\section{Conclusion}\label{sec:concl}

This paper proposes a novel approach to improve the utilization of communication and control resources in \gls{wncs}. 
To this end, a new notion of stability, referred to as tail-based stability, deals with extreme conditions relaxing the classical stability conditions.
Then, joint control and communication are carried out in three stages:
(i) scheduling control systems using the Lyapunov optimization framework,
(ii) predicting missing plant states with \gls{gpr},
and
(iii) deriving the control policy using \gls{rl}.
The results show that by combining the above three mechanisms, the proposed method can outperform designs that rely on different state-of-the-art scheduling, prediction, and controlling methods.
Simultaneously controlling heterogeneous systems and extending the analysis for different propagation environments and multi-antenna systems along with the ablation studies via varying system parameters are left as future directions.

\bibliographystyle{IEEEtran}

\end{document}